\documentclass[10pt,aps,showpacs,twocolumn]{revtex4}
\usepackage[latin9]{inputenc}
\usepackage{amsmath}
\usepackage{amssymb}
\usepackage{graphicx}
\usepackage{esint}
\usepackage{epstopdf}
\makeatletter
\@ifundefined{textcolor}{}
{%
 \definecolor{BLACK}{gray}{0}
 \definecolor{WHITE}{gray}{1}
 \definecolor{RED}{rgb}{1,0,0}
 \definecolor{GREEN}{rgb}{0,1,0}
 \definecolor{BLUE}{rgb}{0,0,1}
 \definecolor{CYAN}{cmyk}{1,0,0,0}
 \definecolor{MAGENTA}{cmyk}{0,1,0,0}
 \definecolor{YELLOW}{cmyk}{0,0,1,0}
}

\usepackage{braket}

\makeatother

\begin{document}

\title{Violation of Cauchy-Schwarz inequalities by spontaneous Hawking radiation
in resonant boson structures}

\author{J. R. M. de Nova, F. Sols and I. Zapata}

\affiliation{Departamento de Física de Materiales, Universidad Complutense de
Madrid, E-28040 Madrid, Spain}


\date{\today}
\begin{abstract}
The violation of a classical Cauchy-Schwarz (CS) inequality is identified
as an unequivocal signature of spontaneous Hawking radiation in sonic
black holes. This violation can be particularly large near the peaks
in the radiation spectrum emitted from a resonant boson structure
forming a sonic horizon. As a function of the frequency-dependent
Hawking radiation intensity, we analyze the degree of CS violation
and the maximum violation temperature for a double barrier structure
separating two regions of subsonic and supersonic condensate flow.
We also consider the case where the resonant sonic horizon is produced
by a space-dependent contact interaction. In some cases, CS violation
can be observed by direct atom counting in a time-of-flight experiment.
We show that near the conventional zero-frequency peak, the decisive
CS violation cannot occur.
\end{abstract}
\maketitle

\section{Introduction}

The emission of Hawking radiation (HR) from the horizon of a black hole
(BH) is an intriguing prediction of modern physics \cite{hawking1974}.
Due to the extremely low effective temperature, its detection in a
cosmological context is unlikely to be achieved in the foreseeable
future. However, it was noted by Unruh \cite{Unruh1976,Unruh1981}
that HR is an essentially kinematic effect that
could be observed on a laboratory scale at temperatures which, while
still too low, lie within conceivable reach. For a quantum fluid passing
through a sonic horizon (i.e., a subsonic-supersonic interface), it
has been predicted \cite{leonhardt2003,leonhardt2003A,balbinot2008,carusotto2008,macher2009,finazzi2010,coutant2010}
that, even at zero temperature, phonons will be emitted into the subsonic
region. Attempts have been made to observe HR in an accelerated Bose-Einstein
(BE) condensate \cite{lahav2010}. An alternative route may be provided
by a quasistationary horizon, which can be achieved by allowing a
confined large condensate to leak in such a way that the outgoing
beam is dilute and fast enough to be supersonic \cite{zapata2011,larre2012}.

HR is a fundamentally quantum phenomenon that results
from the impossibility of identifying the vacuum of incoming quasiparticles
with that of outgoing quasiparticles \cite{birrell1982}. Specifically,
the incoming vacuum is a squeezed state of outgoing quasiparticles.
In this respect, it has been long recognized in quantum optical contexts
\cite{loudon2000,walls2008} that correlation functions characterizing
the electromagnetic radiation satisfy Cauchy-Schwarz (CS) type inequalities,
which can, however, be violated in the deep quantum regime, particularly by
squeezed light. Thus, violation of CS inequalities is generally regarded
as a conclusive signature of quantum behavior. By contrast, detection
schemes based on the space correlation function \cite{balbinot2008}
show no signal difference between spontaneous and thermal (stimulated)
HR processes \cite{recati2009}.

An additional advantage of the focus on the violation of CS inequalities
is that it permits us to distinguish the specific squeezed character
of HR from the general properties of coherent collective behavior.
Measurements of space correlation functions \cite{balbinot2008},
or phonon \cite{schutzhold2006} or atom \cite{zapata2011} intensity
spectra, would not allow for such a distinction. The reason is that
the same Bogoliubov-de Gennes equations describe two different phenomena:
linearized collective motion and quantum quasiparticle excitation.
Collective motion is imprinted on the coherent (condensed) part of
the wave function, which does not describe the squeezed zero-point
dynamics of Bogoliubov quasiparticles. Importantly, we propose that
spontaneous HR can be tested by a CS violation involving two specific
outgoing channels, one traveling against the flow in the subsonic
region and the other one dragged by the flow on the supersonic side.
We identify this particular CS violation as an unequivocal signature
of spontaneous HR. The same cannot be said about other
CS violations that
are always present in a Bogoliubov vacuum; for example, those associated with the
paired atom modes $k,-k$ in a condensate at rest.
Such CS violations will appear in other correlation functions.

It has recently been noted that HR could be observed more easily in
contexts where the predicted spectrum is not thermal but peaked around
a discrete set of frequencies \cite{finazzi2010,zapata2011}. This
could be the case in a sonic horizon formed by a double-barrier structure
\cite{zapata2011}
lying between the subsonic and supersonic asymptotic regions. Optical
analogs can show similar peaked structures \cite{rubino2012}. The
purpose of this paper is to show that, despite the remaining difficulties,
the violation of CS inequalities in HR is comparatively
much easier to observe near the peaks characteristic of resonant radiation.
The quest for CS violation here proposed complements those approaches
relying on the direct detection of entanglement, as applied to inflationary
cosmology \cite{campo2006}, BHs \cite{martinez2010}, general
relativistic quantum fields \cite{friis2012}, or BH
analogs \cite{hortsmann2010,hostmann2011,carusotto2013}.

\begin{figure*}[tb!]
\includegraphics[width=1\textwidth]{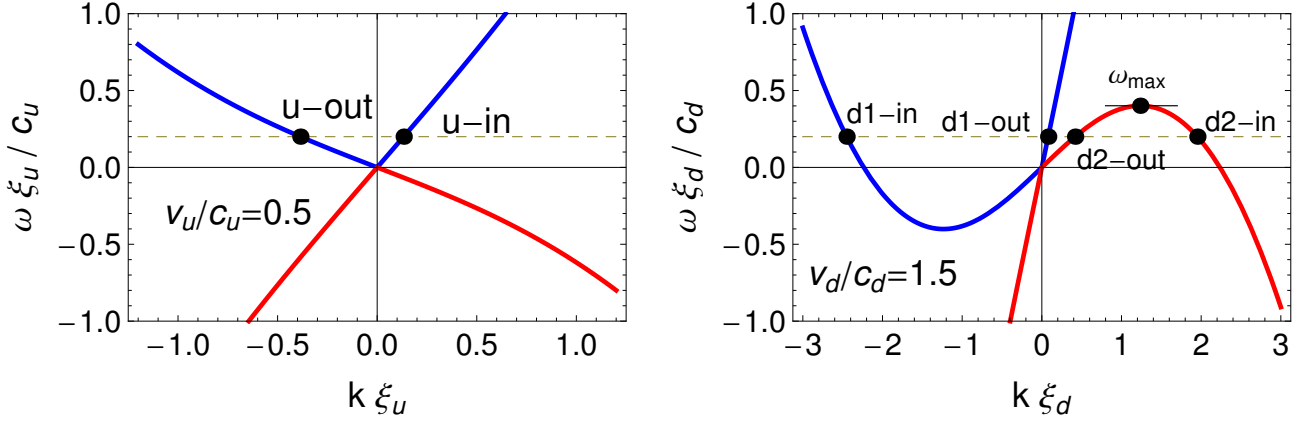}
\caption{(Color online) Quasiparticle dispersion relation on the subsonic (left, upstream) and supersonic
(right, downstream) sides. The blue (red) branches correspond to positive (negative)
normalization. As in Refs. [\onlinecite{recati2009,zapata2011,larre2012}],
$d(u)$ denotes downstream/upstream. Here, $\xi_{u(d)}$ denotes the
asymptotic healing length, $c_{u(d)}$ and $v_{u(d)}$ the sound and
flow velocities, and $\omega_{{\rm max}}$ the frequency above which
no HR can be generated.}
\label{figDispRelation}
\end{figure*}

\section{Cauchy-Schwartz inequalities in Hawking Radiation}

We focus on the properties of the normalized second-order correlation
function, which for light is defined as \cite{loudon2000,walls2008}
\begin{equation}
g_{ij}^{(2)}(\tau)\equiv\frac{\langle\hat{a}_{i}^{\dagger}(0)\hat{a}_{j}^{\dagger}(\tau)\hat{a}_{j}(\tau)\hat{a}_{i}(0)\rangle}{\langle\hat{a}_{i}^{\dagger}(0)\hat{a}_{i}(0)\rangle\langle\hat{a}_{j}^{\dagger}(0)\hat{a}_{j}(0)\rangle}\,,\label{eqCoherenceDef}
\end{equation}
where $\hat{a}_{i}(t)$ is the Heisenberg operator for photon mode
$i$, and the average is quantum-statistical. The correlation function
for classical light is obtained by removing the quantum average and
replacing the Heisenberg operators $\hat{a}_{i}(t)$ by complex numbers.
The following inequalities can be proven for classical light:
\begin{eqnarray}
1 & \leq & g_{ii}^{(2)}(0),\label{eqClassicalIneq-a}\\
g_{ii}^{(2)}(\tau) & \leq & g_{ii}^{(2)}(0),\label{eqClassicalIneq-b}\\
\left[g_{ij}^{(2)}(\tau)\right]^{2} & \leq & g_{ii}^{(2)}(0)g_{jj}^{(2)}(0)\,,\;(i\neq j)\,.\label{eqClassicalIneq-c}
\end{eqnarray}
These inequalities are satisfied not only classically but also by
quantum thermal states at high temperature. They are also satisfied
by chaotic and coherent states.

The violation of any of the above inequalities is a signature of deep
quantum behavior. States violating (\ref{eqClassicalIneq-a}) are
said to show sub-Poissonian statistics. Violation of (\ref{eqClassicalIneq-b})
reflects anti-bunching. Expression (\ref{eqClassicalIneq-c}) is a
CS inequality. States which violate it at $\tau=0$ are said to exhibit
two-mode sub-Poissonian statistics. In general, the proof of (\ref{eqClassicalIneq-c})
requires the system to be described by a positive (Glauber-Sudarshan)
P function. Some quantum states such as two-mode squeezed states may
not satisfy this condition and thus can violate (\ref{eqClassicalIneq-c}).
That could be the case in a collision between two BE condensates \cite{kheruntsyan2012}.


Now we turn our attention to Hawking radiation.
References \onlinecite{recati2009,macher2009,zapata2011,larre2012}
have addressed the existence and possible detection of HR in bosonic
condensates. Figure \ref{figDispRelation} shows the dispersion relation
of the scattering channels following the mode notation of Refs. \onlinecite{recati2009,zapata2011,larre2012},
to which the reader is referred for further details. A particular
type of HR often considered in analog systems is the
emission of $u$-out phonons as originated in the anomalous transmission
from the $d2$-in channel. Hereafter, the operator $\hat{\gamma}_{i-\alpha}$
destroys a quasiparticle in the scattering state characterized by channel $i-\alpha$, with $i=u,d1,d2$
and $\alpha={\rm in,out}$. The dependence of $\hat{\gamma}_{i-\alpha}(\omega)$
on the quasiparticle frequency $\omega$ will often be understood.
As is conventional in finite temperature HR setups, we assume that
averages are taken for a thermal distribution of incoming quasiparticles,
so that $\langle\hat{\gamma}_{i-{\rm in}}^{\dagger}(\omega)\hat{\gamma}_{j-{\rm in}}(\omega')\rangle=n_{i}(\omega)\delta_{ij}\delta(\omega-\omega')$,
where $n_{i}(\omega)=\left[\exp(\hbar\Omega_{i}(\omega)/k_{B}T)-1\right]^{-1}$
and $\Omega_{i}(\omega)$ is the comoving frequency corresponding
to the mode $i$-in at the laboratory frequency $\omega$.

We consider the equal-time second-order correlation function for the
outgoing quasiparticle operators
\begin{equation}
\Gamma_{ij}\equiv\langle\hat{\gamma}_{i-{\rm out}}^{\text{\ensuremath{\dagger}}}\hat{\gamma}_{j-{\rm out}}^{\dagger}\hat{\gamma}_{j-{\rm out}}\hat{\gamma}_{i-{\rm out}}\rangle>0,\label{eqGammaDef}
\end{equation}
and define $\theta_{ij}\equiv\Gamma_{ij}/
\sqrt{\Gamma_{ii}\Gamma_{jj}},\,\Theta_{ij}\equiv\Gamma_{ij}-\sqrt{\Gamma_{ii}\Gamma_{jj}}$ \cite{comm1},
noting that the CS inequality (\ref{eqClassicalIneq-c})
is violated if and only if
\begin{equation}
\theta_{ij}>1\,({\rm or}\,\,\Theta_{ij}>0)\,.\label{eqThetaViolation}
\end{equation}
Thus, we may use $\theta_{ij}$ ($\Theta_{ij}$) as a relative (absolute)
figure of merit to quantify the degree of CS violation.

We define the complex vector $\alpha_{i}^{\dagger}\equiv(\sqrt{n_{u}}S^*_{iu},\sqrt{n_{d1}}S^*_{id1},\sqrt{n_{d2}+1}S^*_{id2})$,
where $S_{ij}$ is the element $ij$ of the scattering matrix $S$
characterizing the transition from $j$-in to $i$-out \cite{comment1}
and obeying the pseudo-unitary condition $S^{\dagger}\eta S=\eta\equiv{\rm diag}(1,1,-1)$.
Specifically,
\begin{equation}\label{eq:inoutmodesrelation}
\left[\begin{array}{c}
\hat{\gamma}_{u-\rm{out}}\\
\hat{\gamma}_{d1-\rm{out}}\\
\hat{\gamma}_{d2-\rm{out}}^{\dagger}
\end{array}\right] = S(\omega)\left[\begin{array}{c}
\hat{\gamma}_{u-\rm{in}}\\
\hat{\gamma}_{d1-\rm{in}}\\
\hat{\gamma}_{d2-\rm{in}}^{\dagger}
\end{array}\right] \, .
\end{equation}

Wick's theorem allows us to write
\begin{eqnarray}
\Gamma_{uu} & = & 2|\braket{\hat{\gamma}_{u-\rm{out}}^{\dagger}\hat{\gamma}_{u-\rm{out}}}|^{2}\nonumber \\
\Gamma_{d2d2} & = & 2|\braket{\hat{\gamma}_{d2-\rm{out}}^{\dagger}\hat{\gamma}_{d2-\rm{out}}}|^{2}\nonumber \\
\Gamma_{ud2} & = & \braket{\hat{\gamma}_{d2-\rm{out}}^{\dagger}\hat{\gamma}_{u-\rm{out}}^{\dagger}}\braket{\hat{\gamma}_{d2-\rm{out}}\hat{\gamma}_{u-\rm{out}}} \nonumber\\&+&\braket{\hat{\gamma}_{u-\rm{out}}^{\dagger}\hat{\gamma}_{u-\rm{out}}}\braket{\hat{\gamma}_
{d2-\rm{out}}^{\dagger}\hat{\gamma}_{d2-\rm{out}}}\,\label{eq:phononGWick},
\end{eqnarray}
which leads to
\begin{eqnarray}
\Gamma_{uu} & = & 2|\alpha_{u}|^{4}\nonumber \\
\Gamma_{d2d2} & = & 2(|\alpha_{d2}|^{2}-1)^{2}\nonumber \\
\Gamma_{ud2} & = & |\alpha_{u}^{\dagger}\cdot\alpha_{d2}|^{2}+|\alpha_{u}|^{2}(|\alpha_{d2}|^{2}-1)\,.\label{eqEtaExpr}
\end{eqnarray}
Making use of (\ref{eqGammaDef}) and (\ref{eqEtaExpr}), the CS inequality (\ref{eqClassicalIneq-c})
for outgoing quasiparticles can be rewritten as
\begin{equation}
|\alpha_{u}^{\dagger}\cdot\alpha_{d2}|^{2}\leq|\alpha_{u}|^{2}\left(|\alpha_{d2}|^{2}-1\right)\,,\label{eqFiniteTempCSViol-JR}
\end{equation}
which, due to the $-1$ term within the bracket, can be violated some
times. Interestingly, the very possibility of violating the CS inequality
(\ref{eqClassicalIneq-c}) is a direct consequence of the anomalous
character of the scattering process $d2$-in $\rightarrow$ $u$-out,
because the $u(d2)$ channel has positive (negative) normalization. In
fact, for the (normal) conversion $d1\leftrightarrow u$, we obtain
that (\ref{eqClassicalIneq-c}) amounts to $|\alpha_{u}^{\dagger}\cdot\alpha_{d1}|^{2}\leq|\alpha_{u}|^{2}|\alpha_{d1}|^{2}$,
which is always satisfied.

The inequality (\ref{eqFiniteTempCSViol-JR}) and pseudo-unitarity lead,
after a lengthy calculation, to the equivalent relation
\begin{eqnarray}
 &  & |S_{ud2}|^{2}(1+n_{u}+n_{d1}+n_{d2})\leq\nonumber \\
 &  & |S_{d1u}|^{2}n_{d1}n_{d2}+|S_{d1d1}|^{2}n_{u}n_{d2}+|S_{d1d2}|^{2}n_{u}n_{d1}\nonumber \\
 &  & +|S_{d2d1}|^{2}n_{u}+|S_{d2u}|^{2}n_{d1}\,.\label{eqFiniteTempCSViol}
\end{eqnarray}
A similar inequality can be derived for the other anomalous process,
$d1\leftrightarrow d2$ (Andreev reflection \cite{zapata2009}), by
interchanging $u$ and $d1$ in (\ref{eqFiniteTempCSViol}).

In the conventional ($\omega$ = 0) peak of the Hawking spectrum $|S_{ud2}(\omega)|^{2}$,
the scattering matrix elements diverge as $|S_{ij}(\omega)|\sim1/\sqrt{\omega}$
with $i$ arbitrary and $j=d1,d2$ . By contrast, $S_{iu}(\omega)$
in the same limit ($\omega\rightarrow0^+$) saturates to a nonzero constant
(the asymptotic behavior of the $S-$matrix coefficients is discussed in Appendix
\ref{sec:SMatrixBehav}). On the other hand, the only occupation factor
which diverges is $n_{u}(\omega)\sim1/\omega$, because $\Omega_{u}(\omega)$
is the only comoving frequency that vanishes for small $\omega$.
From pseudo-unitarity it follows that $|S_{ud2}|^{2}-|S_{d2d1}|^{2}=|S_{d2u}|^{2}-|S_{d1d2}|^{2}$.
We conclude that (\ref{eqFiniteTempCSViol}) cannot be violated in
the $\omega\rightarrow0^+$ region. This argument relies solely on the
presence of a uniform condensate flow connecting the subsonic and supersonic
asymptotic regions, and not on other details of the scattering structure.

It can also be proven that there is no violation
in the $\omega\rightarrow\omega^-_{\rm max}$ region, where $\theta_{ud2}=1/2+O\left(\sqrt{\omega_{\rm max}-\omega}\right)$.
This leaves us with only the central frequency region
in the quest for CS violation. We know of no structure other than resonant BH \cite{zapata2011},
which is able to display peaks in that central region; see Sec. \ref{sec:CSViolationResonanteStructures}
for more details.

The inequality (\ref{eqFiniteTempCSViol}) is manifestly violated
at temperature $T=0$ and $\omega\neq0$ provided $S_{ud2}\neq0$,
which further reflects the direct link between CS violation and HR. In this limit, the condition (\ref{eqThetaViolation})
is equivalent to
\begin{equation}
\theta_{ud2}=\frac{|S_{d2d2}|^{2}-1/2}{|S_{d2d2}|^{2}-1}>1\,,\,\,\Theta_{ud2}=|S_{ud2}|^{2}>0\,,\label{eqZeroTempCSViol}
\end{equation}
where pseudo-unitarity has again been invoked.
The condition (\ref{eqZeroTempCSViol})
is guaranteed to be satisfied whenever $S_{ud2}\neq0$ (in that case, pseudo-unitarity requires $|S_{d2d2}|>1$).
Finally, we note that $\theta_{d1d2}=\theta_{ud2}$
and $\Theta_{d1d2}=|S_{d1d2}|^{2}$ at zero temperature.

A device producing HR works like a nondegenerate parametric amplifier,
which is known to generate squeezing from vacuum. However, sources
other than vacuum may also generate squeezing and ultimately CS violation
\cite{walls2008}. In particular, the absolute amount of CS violation
for a given frequency often increases initially as the temperature
is raised from zero, eventually reaching a maximum and decreasing
to zero at high temperatures, as can be guessed from Eq. (\ref{eqFiniteTempCSViol}).
Therefore, one may wonder whether the CS violation here contemplated
would provide conclusive evidence of spontaneous (zero-point) HR radiation.
The answer is yes, as we argue below.

Wick's theorem can again be invoked to write $\Gamma_{ij}$ in terms
of first-order correlation functions such as $\langle\hat{\gamma}_{i-{\rm out}}^{\dagger}\hat{\gamma}_{j-{\rm out}}\rangle$ or $\langle\hat{\gamma}_{i-{\rm out}}\hat{\gamma}_{j-{\rm out}}\rangle$
[see Eq. (\ref{eq:phononGWick})],
here generically referred to as $\rho_{ij}$ \cite{comm2}.
The thermal contribution
is $\rho_{ij}^{{\rm th}}\equiv\rho_{ij}-\rho_{ij}^{0}$, with $\rho_{ij}^{0}$
the zero-temperature value (average over incoming vacuum). If one
neglects the zero-point contributions by approximating $\rho_{ij}\simeq\rho_{ij}^{{\rm th}}$,
then one arrives at a modified version of (\ref{eqFiniteTempCSViol})
where only the terms quadratic in the $n_{i}$'s survive. The resulting
inequality is always satisfied. We conclude that CS violation requires
vacuum fluctuations.

\begin{figure*}[!tb]
\includegraphics[width=1.05\textwidth]{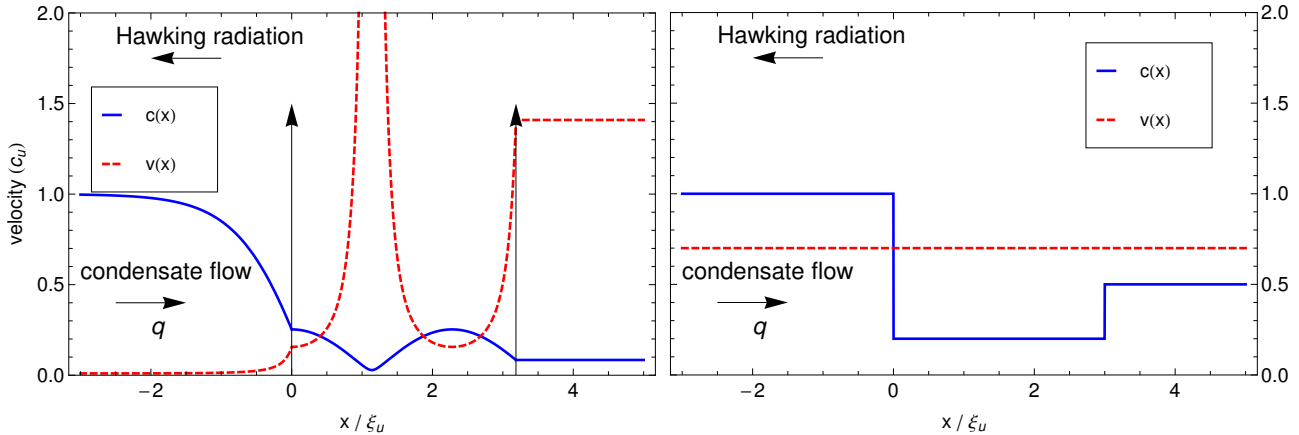}
\caption{(Color online) Scheme of the double-barrier configuration (left) and the configuration
with two discontinuities in the otherwise flat density configuration
(right).}
\label{fig:ResonantScheme}
\end{figure*}

The upshot of this discussion is that the detection of a CS inequality
violation can be regarded as a smoking gun
for the presence of spontaneous HR.
In other words, the spontaneous signal is the
cause of the quantum behavior as reflected in the violation of the classical inequalities.
This contrasts with other
schemes in which the spontaneous signal cannot be distinguished unambiguously
from the stimulated (or thermal) signal, which has, in this sense, a classical
behavior and thus would never originate the CS violation by itself.

We may define the violation temperature $T_{v}$ as the highest temperature at which (\ref{eqFiniteTempCSViol})
is violated {[}(\ref{eqThetaViolation}) is satisfied{]} and try to
identify some trends in its behavior. We note that, for a double-barrier structure \cite{zapata2011},
if the upstream
current is small (kinetic contribution to total chemical potential
on subsonic side is small), then for
$\omega\gtrsim\omega_{{\rm max}}/2$ the comoving frequencies of the
various channels are comparable to the comoving chemical potential on
the subsonic side, $\mu$.

If the conversion from negative to positive normalization is weak,
then $|S_{d2d2}|$ is close to unity and from pseudo-unitarity it
can be proven that all the $S$-matrix elements appearing in (\ref{eqFiniteTempCSViol})
are small except for the relation $|S_{d1u}|^{2}+|S_{d1d1}|^{2}\simeq1$.
If $\ell\equiv-\log|S_{ud2}|\gg1$ we obtain a low violation temperature,
$k_B T_{v}\sim\mu/\ell\ll\mu$.

Conversely, also within the case of a double-barrier structure and for low currents,
we find numerically that for the highest nonzero-frequency HR peaks (where $|S_{ud2}|\!\!\gg\!\!1$)
the approximate relation
$|S_{ud2}|\simeq|S_{d2u}|\gg|S_{d2d1}|$
applies in the vast majority of cases.
At the same time, $|S_{ud2}|\gg|S_{d1j}|$ for all $j$, which implies
that the violation temperature satisfies $k_BT_{v}\sim\mu|S_{ud2}|^{2}/S_{d1}^{2}\gg\mu$,
where $S_{d1}\equiv\max_{j}\{|S_{d1j}|\}$. Thus, $T_{v}(\omega)$
is expected to be large near the peak frequency $\omega_{0}$. Non-resonant
structures can also show CS violation at nonzero temperatures. However,
even if the relative violation (as measured by $\theta_{ud2}$) is
significant, the absolute amount of violation (as measured by $\Theta_{ud2}$)
turns out to be negligible, as discussed later.


Another important advantage of resonant peaks is that, at their relatively
high frequency, the phononic signal is approximately proportional
to the atomic signal, the latter being directly measurable in a time-of-flight
(TOF) experiment. In Appendix \ref{app:Atom2PhononTOF} technical
details are given for the definition of atomic signals and their relation to
phonon measurements. Assuming that the subsonic and supersonic TOF
signals can be experimentally detached, near the peak at $\omega=\omega_{0}$,
and neglecting finite-size effects, the atom operators can be approximated
as:
\begin{eqnarray}
\hat{c}_{u}(p_{u}\equiv q_{u}+k_{u-{\rm out}}) & \sim & \hat{\gamma}_{u-{\rm out}}(\omega)\label{eqTOFPhononU}\\
\hat{c}_{d}(p_{d1}\equiv q_{d}+k_{d1-{\rm out}}) & \sim & \hat{\gamma}_{d1-{\rm out}}(\omega)\label{eqTOFPhononD1}\\
\hat{c}_{d}(p_{d2}\equiv q_{d}-k_{d2-{\rm out}}) & \sim & \hat{\gamma}_{d2-{\rm out}}(\omega)\,,\label{eqTOFPhononD2}
\end{eqnarray}
where $\hat{c}_{u/d}(k)$ annihilates an atom of momentum $k$ on
side $u/d$. Here, $k_{i-{\rm out}}$ is the comoving quasiparticle
momentum at the laboratory frequency $\omega$ in the $i-{\rm out}$ channel,
and $q_{u/d}$ is the condensate momentum per atom on each side. The
proportionality factors not shown in (\ref{eqTOFPhononU})-(\ref{eqTOFPhononD2})
cancel in the CS violation condition (\ref{eqThetaViolation}) for
$\theta_{ud2}$ and $\theta_{d1d2}$.


For example, the approximations (\ref{eqTOFPhononU}),(\ref{eqTOFPhononD2}) apply whenever $|S_{ud2}(\omega_{0})|^{2}\!\gg\!1$, because the Hawking signal has to stand out above that of the depletion cloud \cite{zapata2011}. For completeness, we note that (\ref{eqTOFPhononD1}) and (\ref{eqTOFPhononD2}) apply if $|S_{d1d2}(\omega_{0})|^{2}\!\gg\!1$.


From (\ref{eqZeroTempCSViol}) and the pseudo-unitarity relation $|S_{ud2}|^{2}+|S_{d1d2}|^{2}+1=|S_{d2d2}|^{2}$,
it may appear that a shortcoming of a large peak is its small relative
degree of CS violation, as reflected in $\theta_{ud2}$ lying slightly above unity,
which can be generally inferred from the bound
\begin{equation}
\theta_{ud2}-1\leq\frac{1}{2(|\alpha_{d2}|^2-1)} \, .
\end{equation}

However, this does not imply that the experimental signal is necessarily small.
Quite the opposite, the absolute amount of violation (as measured
by $\Theta_{ud2}$) can be quite large, as (\ref{eqZeroTempCSViol})
directly reveals. A similar analysis can be performed for the other
anomalous process, $d1\leftrightarrow d2$. 

We can define, in analogy to (\ref{eqGammaDef}), the
atomic correlation function
\begin{equation}
G_{ud2}(\omega)\equiv\langle\hat{c}_{u}^{\dagger}(p_{u})\hat{c}_{d}
^{\dagger}(p_{d2})\hat{c}_{d}(p_{d2})\hat{c}_{u}(p_{u})\rangle \, ,\label{eqGDef}
\end{equation}
and similarly for the other $G_{ij}$, where $i,j=u,d1,d2$, and $p_{i}(\omega)$
is defined in (\ref{eqTOFPhononU})-(\ref{eqTOFPhononD2}). It can be
shown (see Appendix \ref{app:Atom2PhononTOF})
that a sufficient condition for (\ref{eqThetaViolation}) is
\begin{equation}
z_{ij}>1\,,\,\, Z_{ij}>0\,,\label{eqZetaViolation}
\end{equation}
where $z_{ij}\equiv G_{ij}/\sqrt{G_{ii}G_{jj}}$ and $Z_{ij}\equiv G_{ij}-\sqrt{G_{ii}G_{jj}}$.

\section{CS violation by Hawking Radiation in resonant structures\label{sec:CSViolationResonanteStructures}}

In order to study CS violation by HR in resonant structures, we focus
on two models. In the first case, a double
delta-barrier potential $V(x)=Z[\delta(x)+\delta(x-d)]$ is introduced
separating the subsonic from the supersonic regions, where $d$ is
the distance between the barriers and $Z$ measures their strength.
This structure was already explored in Ref. \onlinecite{zapata2011} and is
schematically depicted in the left panel of Fig. \ref{fig:ResonantScheme}.
Given these parameters, only a discrete number of currents are
compatible with the existence of two asymptotic regions of uniform
density and flow speed, one subsonic and one supersonic.
Even though the speed of sound is only properly defined for sufficiently
flat regions, by extending its definition in terms of the local density
of the condensate to generally nonuniform profiles,
$c(x)=[gn(x)/m]^{1/2}$,
this speed of
sound will cross the flow velocity at least once.
Here, $g$ is the one-dimensional coupling constant, $n(x)$ is the linear atom density,
and $m$ the atom mass.
As expected from the study in Ref. \onlinecite{zapata2011}, the HR spectrum
displays resonant peaks, its number increasing roughly with the barrier separation.
We will argue that these peaks are good candidates to
exhibit CS violations.

\begin{figure}[!htb]
\includegraphics[width=1\columnwidth]{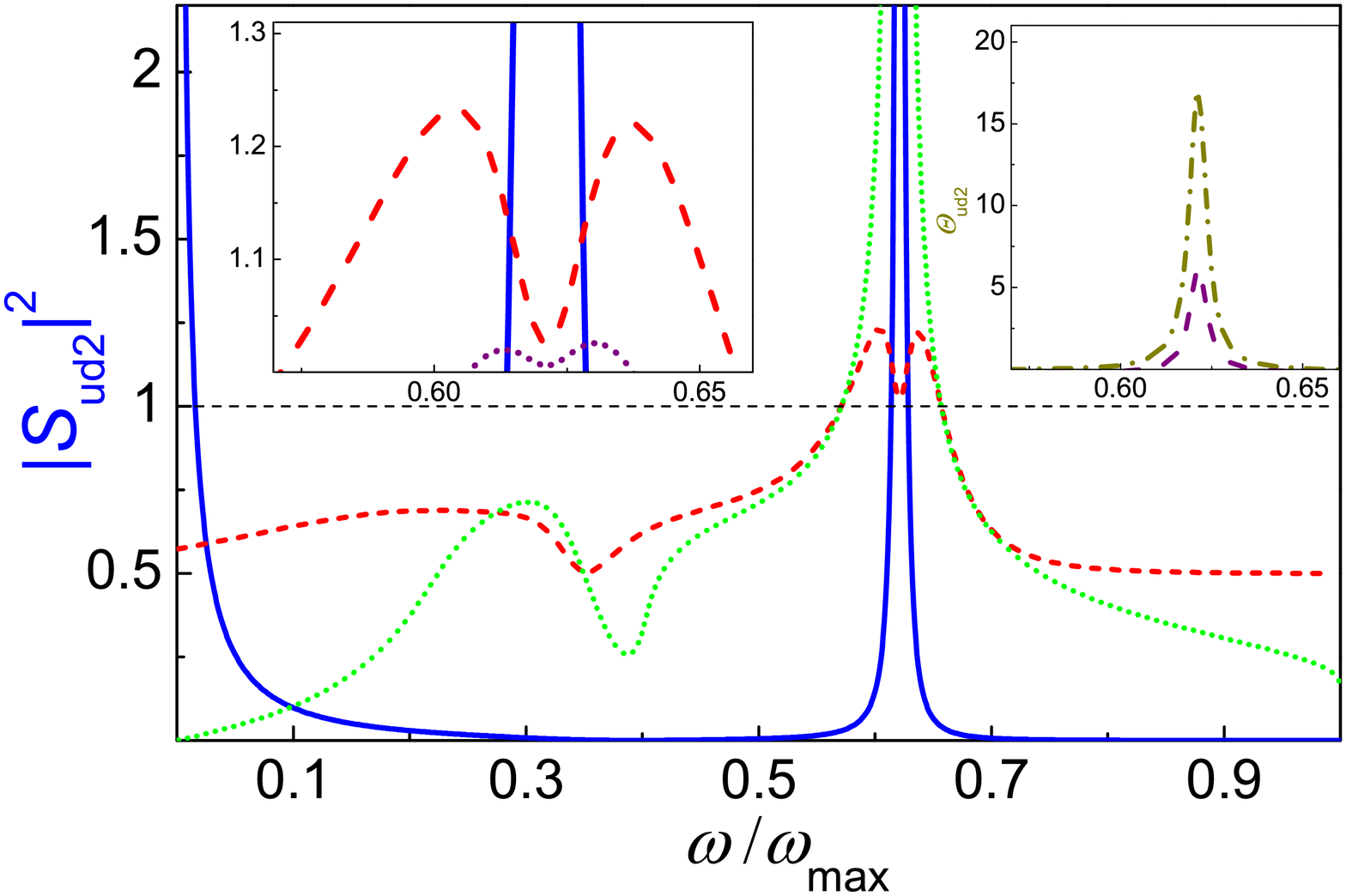}
\caption{(Color online) Hawking radiation for a condensate leaking through a double barrier
structure like that studied in Ref. \onlinecite{zapata2011} and shown in the left
Fig. [\ref{fig:ResonantScheme}]. The
strength of both delta barriers is $Z=2.2\hbar^{2}/m\xi_{u}$. They
are separated by a distance $d=3.62\xi_{u}$. The flow is such that
$q_{u}\xi_{u}=0.01$ and $\omega_{{\rm max}}=0.99\mu/\hbar$. Solid
blue: zero-temperature HR spectrum $|S_{ud2}(\omega)|^{2}$.
Dashed red: $\theta_{ud2}(\omega)$ at temperature $T=\mu/k_{B}$,
where $\mu$ is the comoving chemical potential on the subsonic side.
Dotted green: maximum violation temperature $T_{v}(\omega)$ in units
of $\mu/k_{B}$. Not shown in the figure, $T_{v}(\omega)$ rises up
to $T_{v}(\omega_{0})\simeq21\mu/k_{B}$, where $|S_{ud2}(\omega_{0})|^{2}\simeq8$.
Left inset: Zoom of the peak region, with $T_{v}(\omega)$ removed
and $z_{ud2}(\omega)$ (relative atom CS violation) added (dotted
purple). Right inset: Same as left inset; it shows $\Theta_{ud2}(\omega)$
(dashed-dotted brown) and $Z_{ud2}(\omega)$ (dashed purple), which
measure the absolute amount of CS violation in the phonon and atom
signals. }
\label{fig:GraphCSViol_2DeltaV1}
\end{figure}

The other model considered in this work is a resonant generalization
of a flat profile configuration, first considered in Ref. \onlinecite{recati2009}
and schematically depicted in the right panel of Fig. \ref{fig:ResonantScheme}.
In that scenario, the GP wave function is everywhere the same plane
wave, $\Psi_{0}(x)=\sqrt{n_{0}}e^{iqx}$, with flow speed
$v=\hbar q/m$. Therefore, the condensate density, velocity and current
are all constant along the BH structure. Both the one-dimensional coupling strength
$g\left(x\right)$ and the external potential $V\left(x\right)$
are space dependent and chosen in such a way that $g'(x)n_{0}+V'(x)=0$.
We consider a spatial dependence with three regions within each of which
the sound speed, $c(x)=[g(x)n_{0}/m]^{1/2}$, is constant. Specifically, we take
$c(x<0)=c_{1}$,
$c(0\leq x\leq L)=c_{2}$, $c(x>L)=c_{3}$, with $c_{1}>v>c_{3}$. That is, the left-most (right-most)
region is subsonic (supersonic). The middle region, with speed of sound $c_2$, can be subsonic
of supersonic depending on the externally chosen parameters. This
highly idealized model yields closed formulas for many quantities of interest.

In Fig. \ref{fig:GraphCSViol_2DeltaV1} we plot, for a double delta-barrier
structure, the zero temperature HR spectrum $|S_{ud2}(\omega)|^{2}$,
together with $\theta_{ud2}(\omega)$ (at $k_{B}T=\mu$) and the violation
temperature $T_{v}(\omega)$. The left inset magnifies the peak region
and includes $z_{ud2}(\omega)$, which measures the relative amount
of CS violation in the atomic signal. The right inset shows $\Theta_{ud2}(\omega)$
and $Z_{ud2}(\omega)$, i.e., the absolute amount of CS violation
by the phonon and atom signals. These graphs indicate that the considered structure
is a promising scenario for the unambiguous detection of spontaneous
HR.

%
%

\begin{figure}[!htb]
\includegraphics[width=1\columnwidth]{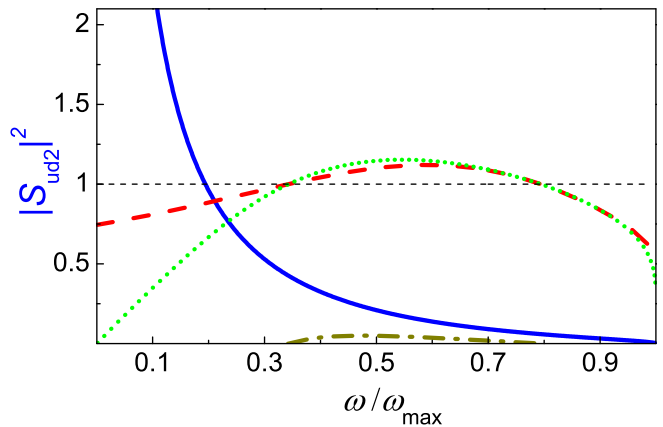}
\caption{(Color online) Same as Fig. \ref{fig:GraphCSViol_2DeltaV1}
for a setup with a single delta barrier of strength $Z=0.62\hbar^{2}/m\xi_{u}$
and a uniform interaction, with flow $q_{u}\xi_{u}=0.3$ and $\omega_{{\rm max}}=0.6\mu/\hbar$.
Atomic CS violation is not found for this setup.}
\label{fig:CSViol_1Delta}
\end{figure}

Figure \ref{fig:CSViol_1Delta} shows the
corresponding curves for a single delta-barrier BH configuration.
The HR spectrum is unstructured, with a single peak at $\omega=0$.
The CS inequality can be violated at the relatively high temperature
$T=\mu/k_{B}$ within a considerable frequency range.
However, the smallness of the absolute phonon violation
signal suggests
that nonresonant structures are not good candidates for the observation
of CS violation. In the cases we have explored, we have not found atomic CS
violation at the temperature $T=\mu /k_B$.

\begin{figure}[!htb]
\includegraphics[width=1\columnwidth]{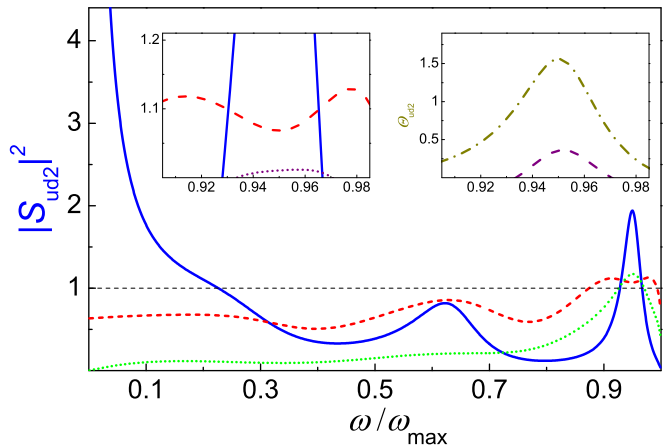}
\caption{(Color online) Same as Fig. \ref{fig:GraphCSViol_2DeltaV1} but for $T=0.6\mu/k_B$ and
for a structure without
barriers but with two sharp variations in the local speed of sound,
which takes the successive values $c_{u},0.43c_{u},0.6c_{u}$,
as depicted in the right Fig. \ref{fig:ResonantScheme}. The
intermediate region has a length $L=26\xi_{u}$. The flow is such that
$q_{u}\xi_{u}=0.95$ and $\omega_{{\rm max}}=0.18\mu/\hbar$. This
is a resonant generalization of the model first studied in Ref. \onlinecite{balbinot2008},
where only one sound-speed discontinuity was considered. }
\label{fig:CSResFlatProfile}
\end{figure}

Figure \ref{fig:CSResFlatProfile} shows the same
curves as those of Fig. \ref{fig:GraphCSViol_2DeltaV1} but for a setup with a flat density
profile such as that depicted in the right Fig. \ref{fig:ResonantScheme}.
The absolute CS violation
is here considerably smaller than in Fig. \ref{fig:GraphCSViol_2DeltaV1}
but could still be observable.

\section{Conclusions}

In conclusion, we find that violation of the Cauchy-Schwartz inequalities in the
HR of strongly peaked spectrum (such as that emitted
by a double-barrier sonic BH) may provide
a convenient route to the unambiguous observation of the zero-point
contribution. In some setups, the violation of CS inequalities is
large enough to be detectable by the direct observation of second-order
correlation functions in an atom time-of-flight experiment. The large absolute
CS violation is absent in nonresonant structures such as that formed
by a single-barrier sonic BH. In particular, the relevant
CS violation cannot occur near the conventional, zero-frequency HR peak universally shown by all one-dimensional black-hole structures.


We thank D. Guery-Odelin, R. Parentani and C. Westbrook for valuable
discussions. Support from MINECO (Spain) through grant FIS2010-21372
and from Comunidad de Madrid through grant MICROSERES-CM (S2009/TIC-1476)
is also acknowledged.

\appendix

\section{Limiting behavior of scattering matrix elements near thresholds\label{sec:SMatrixBehav}}

In this Appendix we study the asymptotic behavior of the amplitudes
of the scattering matrix coefficients, $|S_{ij}\left(\omega\right)|$,
close to the thresholds of the HR spectrum, i.e., in the limits
$\omega\rightarrow0^+$ and $\omega\rightarrow\omega^-_{\rm max}$. In the following, we adopt units $\hbar=m=1$.
In the asymptotic regions where the condensate flow is uniform, the condensate
wave function is of the form $\Psi_0(x)=\sqrt{n_{r}}e^{i(q_{r}x+\alpha_{r})}$,
where $r=u,d$ labels the asymptotic region at $x\rightarrow\text{\ensuremath{\pm\infty}}$.
The asymptotic plane-wave solutions to
the Bogoliubov-de Gennes (BdG) equations at a given frequency $\omega$
(scattering channels) can
be written as

\begin{eqnarray}
s_{a,\omega}\left(x\right) & := & \frac{e^{ik_{a}\left(\omega\right)x}}{\sqrt{2\pi|w_{a}\left(\omega\right)|}}\left[\begin{array}{c}
e^{iq_{r}x}u_{a}(\omega)\\
e^{-iq_{r}x}v_{a}(\omega)
\end{array}\right] \nonumber\\
u_{a}(\omega) & = &C_a
\left\{\frac{k_{a}^{2}\left(\omega\right)}{2}+[\omega-q_{r}k_{a}\left(\omega\right)]\right\}\nonumber \\
v_{a}(\omega) & = & C^{*}_a
\left\{\frac{k_{a}^{2}\left(\omega\right)}{2}-[\omega-q_{r}k_{a}\left(\omega\right)]\right\} \, \nonumber\\
C_a&=&\frac{e^{i\alpha_{r}}}
{
\sqrt{2k_{a}^{2}\left(\omega\right)\left|\omega-q_{r}k_{a}\left(\omega\right)\right|}}. \label{eq:PlaneWaveSpinors-1}
\end{eqnarray}

We also refer to $s_{a,\omega}\left(x\right)$ as the spinor (two-component wave function) of the propagating mode or scattering
channel $a$ at frequency $\omega$.

Here, the mode index $a$ labels the four wave vectors which are solutions of the
Bogoliubov's dispersion relation
\begin{equation}
\left[\omega-q_{r}k_{a}\left(\omega\right)\right]^{2}=
c_{r}^{2}k_{a}^{2}\left(\omega\right)+k_{a}^{4}\left(\omega\right)/4
\label{k-solutions}
\end{equation}
for a given frequency $\omega>0$ in the region of index $r=u,d$, while $w_{a}\left(\omega\right):=\left[dk_{a}\left(\omega\right)/d\omega\right]^{-1}$
is the group velocity of mode $a$ which lives in the asymptotic region $r$. The sound velocities are
$c_r=\sqrt{g_r n_{r}}$, where $g_{r}$ is the corresponding asymptotic value of the coupling constant $g(x)$.
In the case of propagating modes,
the index $a$ is of the form $a=i-\alpha$ (with $i=u,d1,d2$ and $\alpha={\rm in,out}$) and
the normalization is $\intop\mathrm{d}x\, s_{a,\omega}^{\dagger}\left(x\right)\sigma_{z}s_{a,\omega'}\left(x\right)=\pm\delta\left(\omega-\omega'\right)$,
with $\sigma_{z}$ the Pauli matrix. The symbol $\pm$ stands for positive or negative normalization.
In the subsonic region, there are only two propagating modes (labeled $u-{\rm in}$ and $u-{\rm out}$),
both with positive normalization, the other two solutions of (\ref{k-solutions}) describing an evanescent wave and
an (unphysical) exploding one.
In the supersonic region, where four propagating modes exist, the two pairs of modes with positive (negative) normalization are denoted as $d1(d2)$,
each pair consisting of an ``in'' and an ``out'' channel.

In the nonhomogeneous case, the field operator in the BdG approximation can be written as $\hat{\Psi}(x)=\Psi_{0}(x)+\delta\hat{\Psi}(x)$
with \cite{zapata2011}:
\begin{eqnarray}
\delta\hat{\Psi}(x) & = & \int_{0}^{\infty}d\omega\sum_{a={\rm u-\rm{in},d1-\rm{in}}}[u_{a,\omega}(x)\hat{\gamma}_{a}(\omega)\nonumber\\&+&v_{a,\omega}^{*}(x)\hat{\gamma}_{a}^{\dag}(\omega)]
 +  \int_{0}^{\omega_{{\rm max}}}d\omega[u_{{\rm {d2-\rm{in},\omega}}}(x)\hat{\gamma}_{{\rm {d2-\rm{in}}}}^{\dag}(\omega)\nonumber\\&+&v_{{\rm {d2-\rm{in},\omega}}}^{*}(x)\hat{\gamma}_{{\rm {d2-\rm{in}}}}(\omega)]\label{eq:FieldOperator-1} \, .
\end{eqnarray}
A similar expression can be written in terms of the ``out'' states. Here, the spinors $z_{a,\omega}(x):=\left[u_{a,\omega}(x),v_{a,\omega}(x)\right]^{\intercal}$ are
solutions to the BdG equations at a given frequency $\omega$, carrying unit flux in the incoming channels $a=d1,d2,u-{\rm in}$
which characterize them. As expected from scattering states, they
contain $S$-matrix coefficients describing the transition from the incoming to the outgoing modes.
For instance,
\begin{eqnarray}
 z_{d2-\rm{in},\omega}\left(x\rightarrow-\infty\right)&=&S_{ud2}\left(\omega\right)s_{u-\rm{out},\omega}\left(x\right)\\
 z_{d2-\rm{in},\omega}\left(x\rightarrow\infty\right)&=&s_{d2-\rm{in},\omega}\left(x\right)\nonumber\\&+&
 S_{d1d2}\left(\omega\right)s_{d1-\rm{out},\omega}\left(x\right)\nonumber\\&+&S_{d2d2}\left(\omega\right)s_{d2-\rm{out},\omega}\left(x\right) \, .\nonumber
\end{eqnarray}
Similar expressions can be written for the other retarded scattering states.

When $\omega\rightarrow0^+$, both the ``$u-$in'' and the three ``out'' scattering channels have
the property $k_{a}\propto\omega$ (see Fig. \ref{figDispRelation}), where $k_a$ is the momentum of the scattering channel $a$.
From (\ref{eq:PlaneWaveSpinors-1}) we infer that the spinors of these four scattering channels behave as
$\propto z_{0}\left(x\right)/\sqrt{\omega}$ with corrections $\sim\sqrt{\omega}$, where
\begin{equation}
z_0(x) \equiv \left[ \Psi_0(x), -\Psi_0^*(x) \right]^{\intercal} \, , \label{zero-mode-spinor}
\end{equation}
is but the zero-mode spinor which solves the BdG equations for $\omega=0$.
The other spinors do not
show this behavior for vanishing $\omega$. Using Cramer's rule to match the left and right solutions, one finds
that $\left|S_{ij}\left(\omega\right)\right|\propto1/\sqrt{\omega}$
for $i,j=d1,d2$. To obtain the behavior of the other scattering matrix
coefficients, it is crucial to note the zero-mode character of $z_{0}\left(x\right)$.
One must also use the property that, in the scattering region
(that between the asymptotic subsonic and supersonic regions), at
least one linear combination of the four solutions tends to the zero-mode
when $\omega\rightarrow0^+$, within corrections to the spinor $z_0(x)$ of order
$\omega$. This is guaranteed because in the BdG equations $\omega$
enters as a nonsingular parameter, i.e., a parameter that is not multiplying
the highest order derivative of the differential equation. Here we have used a theorem
(sometimes referred to as Poincaré's theorem) on the analytic dependence
of a solution to a differential equation on its parameters
(see Refs. \onlinecite{arnold1992,kaltchev2011}).

Using these results, the same type of reasoning as before leads to $\left|S_{ui}\left(\omega\right)\right|\propto1/\sqrt{\omega}$
for $i=d1,d2$ and to a nondivergent amplitude for the remaining matrix elements ($S_{iu}$ with $i=d1,d2$).

In the opposite threshold, $\omega\rightarrow\omega^-_{\rm max}$, the only
spinors that show irregular behavior are those involving the $d2$ mode. They become
proportional to each other because $k_{d2-\rm{in/out}}\left(\omega\right)=k_{d2}(\omega_{\rm max})\pm O(\sqrt{\omega_{\rm max}-\omega})$
and hence, the group velocity vanishes as $\sim\sqrt{\omega_{\rm max}-\omega}$.
The corresponding spinors behave as $s_{d1-{\rm in/out},\omega}(x) \propto s_{\omega_{\rm max}}\left(x\right) \left(\omega_{\rm max}-\omega\right)^{-1/4}+O\left(\omega_{\rm max}-\omega\right)^{1/4}$
where $s_{\omega_{\rm max}}\left(x\right)$ is the value of the spinor
evaluated at $\omega_{\rm max}$. Using again Cramer's rule, it can be
shown that $S_{ud2}\left(\omega\right),S_{d1d2}\left(\omega\right),S_{d2u}\left(\omega\right),S_{d2d1}\left(\omega\right)\propto\left(\omega_{\rm max}-\omega\right)^{1/4}$,
while $S_{d2d2}\left(\omega\right)=-1+O(\sqrt{\omega_{\rm max}-\omega})$
.

\section{Correlation between phonon and atomic time-of-flight signals.\label{app:Atom2PhononTOF}}

For a given frequency $\omega>0$, we may define $p_{u}(\omega)\equiv q_{u}+k_{u-{\rm out}}(\omega)$,
$p_{d2}(\omega)\equiv q_{d}-k_{d2-{\rm out}}(\omega)$ [see Eqs. (\ref{eqTOFPhononU}),(\ref{eqTOFPhononD2})
and Appendix \ref{sec:SMatrixBehav} for the mode notation] as the
corresponding atomic wave vectors. In a time-of-flight (TOF) experiment,
we require the atomic wave vectors from the upstream (downstream)
region to be negative (positive), i.e., we just consider frequencies where
$p_{u}(\omega)<0$ \cite{comm3}.
In particular, this must
be valid at the peak frequency $\omega=\omega_{0}$, i.e., one must have
$p_{u}(\omega_{0})<0$.
If all these conditions are fulfilled, then there is
no contribution from the condensate wave function to the atomic signals.

\begin{figure*}[!tb]
\begin{tabular}{@{}cc@{}}
    \includegraphics[width=\columnwidth]{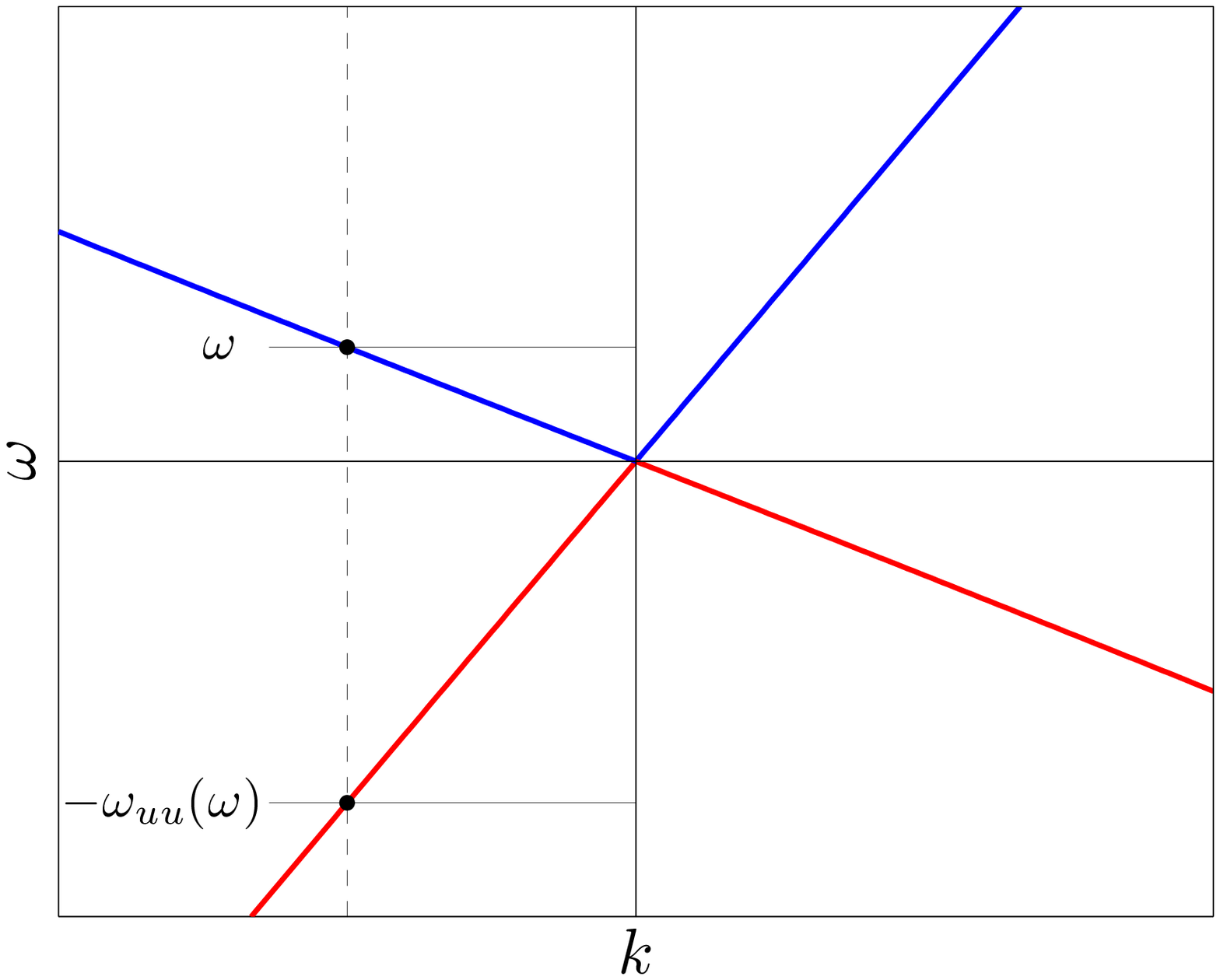} &
    \includegraphics[width=0.958\columnwidth]{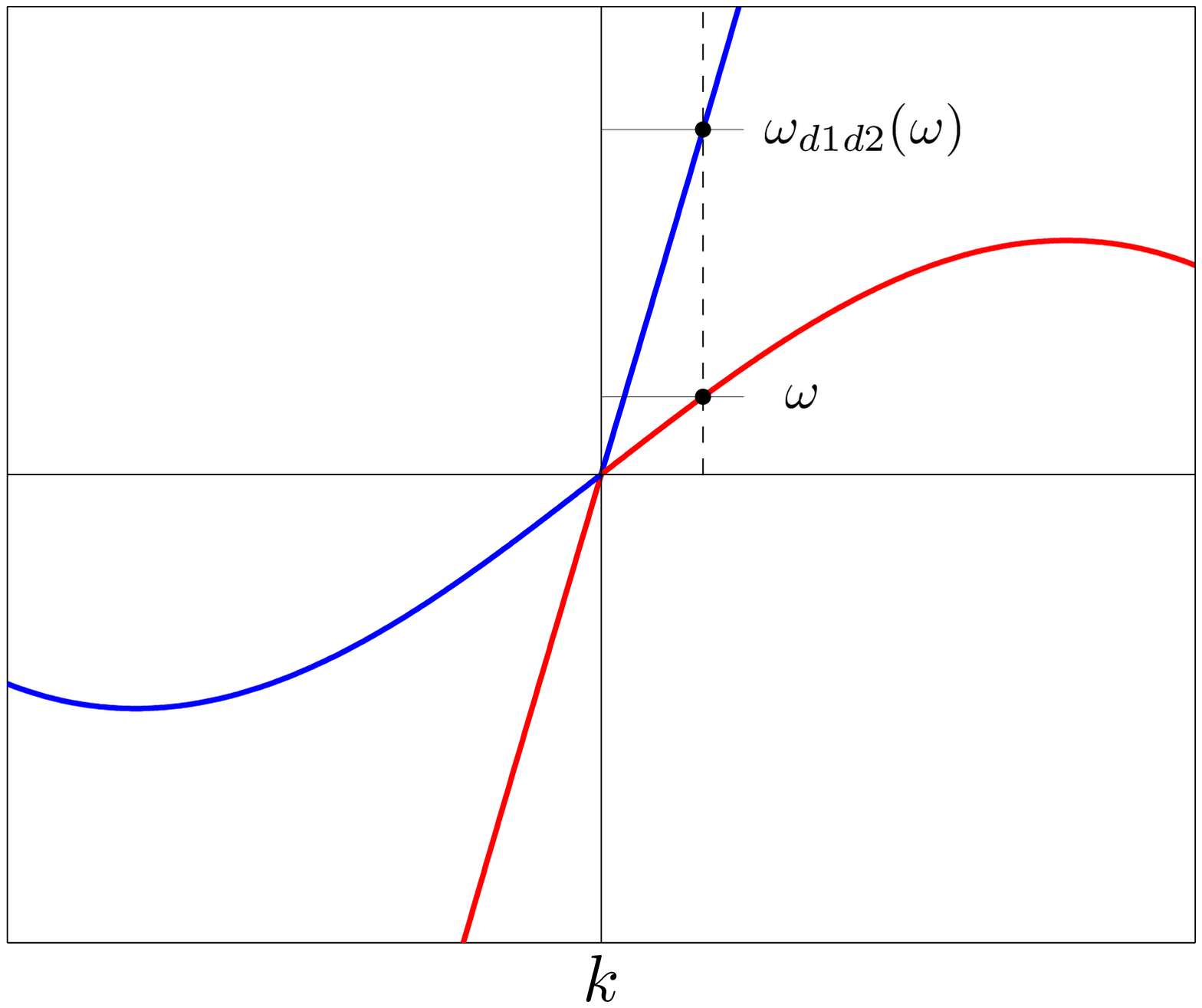} \\
\end{tabular}
\caption{(Color online) Scheme that shows graphically the values of $\omega_{d1d2}(\omega)$ and $-\omega_{uu}(\omega)$.
See Fig. \ref{figDispRelation} for mode notation.}
\label{fig:DepletionOmegas}
\end{figure*}

We define the atom destruction operators as
\begin{align}
\nonumber\hat{c}_{u}\left( p \right) & \equiv\int\mathrm{d}x\, f_{u}^{*}\left(x\right)
e^{-i  p x}
\delta\hat{\Psi}(x)\\
\hat{c}_{d}\left(p \right) & \equiv\int\mathrm{d}x\, f_{d}^{*}\left(x\right)e^{-ipx}\delta\hat{\Psi}(x),
\end{align}
where $f_{r}\left(x\right)$ are normalized functions ($\int\mathrm{d}x|f_{r}(x)|^{2}=1$)
localized only in the asymptotic $r=u,d$ regions and
their Fourier transforms are $f_{r}(k)=\frac{1}{\sqrt{2\pi}}\int\mathrm{d}x~e^{-ikx}f_{r}\left(x\right)$. For simplicity, we
assume that they are of the form
\begin{widetext}
\begin{equation}
f_{r}(x)=\frac{1}{\sqrt{L_{r}}}f\left(\frac{x-x_{r}}{L_{r}}\right) \, ,
\end{equation}
where $f$ is a symmetric and real dimensionless function, $x_{r}$ is the point
at which the asymptotic region $r$ is assumed to be
centered, and $L_{r}$ is the typical size of that region.
We assume that both such lengths are much greater than the corresponding
healing lengths ($L_{r}\gg\xi_{r}$),  so that $f_{r}(k)$ are well peaked at zero momentum. Also, we suppose that
the scattering region is much smaller than the asymptotic regions.
Then, the atomic operators can be approximated as
\begin{eqnarray}
\nonumber\hat{c}_{u}\left(p_{u}\left(\omega\right)\right) & \simeq & \int\mathrm{d}\omega'~f^{*}_u(k_{u-\rm{out}}(\omega')-k_{u-\rm{out}}(\omega))\frac{u_{u-\rm{out}}(\omega')}
{\left|w_{u-\rm{out}}\left(\omega'\right)\right|^{1/2}}
\hat{\gamma}_{u-\rm{out}}(\omega')\\
 \nonumber& + & f^{*}_u(-k_{u-\rm{in}}(\omega')-k_{u-\rm{out}}(\omega))
\frac{v_{u-\rm{in}}^{*}(\omega')}
{|w_{u-\rm{in}}\left(\omega'\right)|^{1/2}}
\hat{\gamma}_{u-\rm{in}}^{\dag}(\omega')\\
\nonumber\hat{c}_{d}\left(p_{d2}\left(\omega\right)\right) & \simeq & \int_{0}^{\omega_{\rm max}}\mathrm{d}\omega'~f^{*}_d(k_{d2-\rm{out}}(\omega)-k_{d2-\rm{out}}(\omega'))
\frac{v_{d2-\rm{out}}^{*}(\omega')}
{|w_{d2-\rm{out}}\left(\omega'\right)|^{1/2}}
\hat{\gamma}_{d2-\rm{out}}(\omega')\\
 & + & \int~\mathrm{d}\omega'~f^{*}_d(k_{d2-\rm{out}}(\omega)-k_{d1-\rm{out}}(\omega'))
 \frac{v_{d1-\rm{out}}^{*}(\omega')}
 {|w_{d1-\rm{out}}\left(\omega'\right)|^{1/2}}
 \hat{\gamma}_{d1-\rm{out}}^{\dag}(\omega'),
\label{c-operator}
\end{eqnarray}

From these two expressions, it can be shown that,
if $|S_{ud2}(\omega_{0})|^{2}\!\gg\!1$,
then the main contribution to the atomic operators near the peak
frequency ($\omega=\omega_{0}$) comes from their respective phononic
counterparts, i.e., $\hat{c}_{u}\left(p_{u}\left(\omega_{0}\right)\right)\sim\hat{\gamma}_{u-\rm{out}}(\omega_{0})$
and $\hat{c}_{d}\left(p_{d2}\left(\omega_{0}\right)\right)\sim\hat{\gamma}_{d2-\rm{out}}(\omega_{0})$,
as expressed in the main text [see Eqs. (\ref{eqTOFPhononU}) and (\ref{eqTOFPhononD2})]. These approximate operator relations
are physically appealing but are not used in our calculations.

From Eq. (\ref{c-operator}) and its preceding discussion, we can calculate the expectation values that
are necessary to compute the second-order correlation functions, namely,
\begin{eqnarray}
\braket{\hat{c}_{u}^{\dagger}\left(p_{u}\left(\omega\right)\right)\hat{c}_{u}\left(p_{u}\left(\omega\right)\right)} & = & |u_{u-\rm{out}}(\omega)|^{2}|\alpha_{u}(\omega)|^{2}\nonumber \\
 & + & |v_{u-\rm{in}}(\omega_{uu}(\omega))|^{2}[1+n_{u}(\omega_{uu}(\omega))]\nonumber \\
\braket{\hat{c}_{d}^{\dagger}\left(p_{d2}\left(\omega\right)\right)\hat{c}_{d}\left(p_{d2}\left(\omega\right)\right)} & = & |v_{d2-\rm{out}}(\omega)|^{2}[|\alpha_{d2}(\omega)|^{2}-1]\nonumber \\
 & + & |v_{d1-\rm{out}}(\omega_{d1d2}(\omega))|^{2}[|\alpha_{d1}(\omega_{d1d2}(\omega))|^{2}+1]\nonumber \\
\braket{\hat{c}_{d}\left(p_{d2}\left(\omega\right)\right)\hat{c}_{u}\left(p_{u}\left(\omega\right)\right)} & = & [\alpha_{d2}^{\dag}(\omega)\cdot \alpha_{u}(\omega)] u_{u-\rm{out}}(\omega)v_{d2-\rm{out}}^{*}(\omega)F(\omega),\label{eq:2orderterm}
\end{eqnarray}
where $\omega_{d1d2}(\omega)=\omega_{d1}(k_{d2-\rm{out}}(\omega))$ and
$\omega_{uu}(\omega)=\omega_{u}(-k_{u-\rm{out}}(\omega))$, with $\omega_{i}(k)$
the dispersion relation of scattering channel $i$ (see Fig. \ref{fig:DepletionOmegas}) . The overlap function is
\begin{eqnarray}
F(\omega) & = & \int\mathrm{d}k~f_{u}^{*}\left(\frac{k}{\rho(\omega)}\right)f_{d}^{*}\left(k\rho(\omega)\right)\nonumber \\
 & = & \frac{1}{\sqrt{L_uL_d}}\int\mathrm{d}x\, f\left(\frac{\rho\left(\omega\right)x+x_{u}}{L_{u}}\right)f\left(\frac{\rho\left(\omega\right)^{-1}x-x_{d}}{L_{d}}\right)\nonumber \, , \\
\rho(\omega) & \equiv &
\left|\frac{w_{u-\rm{out}}(\omega)}{w_{d2-\rm{out}}(\omega)}\right|^{1/2} \, .
\end{eqnarray}
\end{widetext}

As we are assuming that the size of the asymptotic regions are
much larger than the size of the scattering region, we can take
$x_{d}/L_{d}=-x_{u}/L_{u}\simeq 1/2$. The integral $F\left(\omega\right)$ can be interpreted as
the scalar product of two normalized functions, so $F(\omega)\leq1$.
This inequality saturates for $\omega$ such that $\rho(\omega)=\sqrt{L_{u}/L_{d}}$.
In particular, we can choose the size of both asymptotic regions to be such that the saturation is achieved at the peak frequency, i.e., such that $\rho(\omega_{0})=\sqrt{L_{u}/L_{d}}$
and therefore $F(\omega_{0})=1$. The criterion $\rho(\omega)=\sqrt{L_{u}/L_{d}}$
has a straightforward physical interpretation. The phonon operators
are evaluated in the frequency domain. The frequency resolution near a given frequency for both wave packets is given by $\Delta\omega_{u-\rm{out}/d2-\rm{out}}=|w_{u-\rm{out}/d2-\rm{out}}\left(\omega\right)|\Delta k_{u/d}$.
The resolution in momentum space is $\Delta k_{r}\sim1/L_{r}$.
We are correlating the phonon operators of $u-\rm{out}$ and $d2-\rm{out}$
modes in the vicinity of a given frequency, so the maximum correlation
is achieved when $\Delta\omega_{u-\rm{out}}=\Delta\omega_{d2-\rm{out}}$ and
this implies the criterion mentioned above.

Then, using Wick's theorem, the correlation functions can be written
in terms of these expectation values as:
\begin{eqnarray}
\label{eq:atomicGWick}
G_{uu}(\omega) & = & \braket{\hat{c}_{u}^{\dagger}\left(p_{u}\right)\hat{c}^{\dagger}_{u}\left(p_{u}\right)\hat{c}_{u}\left(p_{u}\right)\hat{c}_{u}\left(p_{u}\right)}\nonumber\\&=&2|\braket{\hat{c}_{u}^{\dagger}\left(p_{u}\right)\hat{c}_{u}\left(p_{u}\right)}|^{2}\nonumber \\
G_{d2d2}(\omega) & = & \braket{\hat{c}_{d}^{\dagger}\left(p_{d2}\right)\hat{c}_{d}^{\dagger}\left(p_{d2}\right)\hat{c}_{d}\left(p_{d2}\right)\hat{c}_{d}\left(p_{d2}\right)}\nonumber\\&=&2|\braket{\hat{c}_{d}^{\dagger}\left(p_{d2}\right)\hat{c}_{d}\left(p_{d2}\right)}|^{2}\\
\nonumber G_{ud2}(\omega) & = & \braket{\hat{c}_{d}^{\dagger}\left(p_{d2}\right)\hat{c}_{u}^{\dagger}\left(p_{u}\right)\hat{c}_{d}\left(p_{d2}\right)\hat{c}_{u}\left(p_{u}\right)}\nonumber\\&=&\braket{\hat{c}_{d}^{\dagger}\left(p_{d2}\right)\hat{c}_{u}^{\dagger}\left(p_{u}\right)}
\braket{\hat{c}_{d}\left(p_{d2}\right)\hat{c}_{u}\left(p_{u}\right)}\nonumber\\ &+&\braket{\hat{c}_{d}^{\dagger}\left(p_{d2}\right)\hat{c}_{d}\left(p_{d2}\right)}\braket{\hat{c}_{u}^{\dagger}\left(p_{u}\right)\hat{c}_{u}\left(p_{u}\right)}\nonumber \, .
\end{eqnarray}
In this language, the violation of the atom CS inequality reads $G_{ud2}>\sqrt{G_{uu}G_{d2d2}}$. Now, comparing (\ref{eq:2orderterm}) and (\ref{eq:atomicGWick}) with (\ref{eq:phononGWick}) and (\ref{eqEtaExpr}), one can infer that:
\begin{equation}
\frac{G_{ud2}}{\sqrt{G_{uu}G_{d2d2}}}<\frac{\Gamma_{ud2}}{\sqrt{\Gamma_{uu}\Gamma_{d2d2}}}
\end{equation}
i.e.
\begin{equation}
z_{ud2}<\theta_{ud2}\label{eq:atomicphononrelativeviolations} \, .
\end{equation}

A similar proof can be invoked for the other index pairs $i \neq j$, the case $i=j$ being trivial. We conclude that the violation of the atom CS inequality is a sufficient condition for the same violation in the phonon signal.



\begin{thebibliography}{10}
\bibitem{comm1} Unless otherwise stated, we are not showing trivial Dirac-delta factors
which ensure equal frequency to all phonons that participate in any
correlation function.
\bibitem{comment1} Note that, because of the condensate flow, in
general $S_{ij}\neq S_{ji}$.
\bibitem{comm2} Note that the second type of quantum thermal average is nonzero only if $i$ has a positive normalization
and $j$ a negative one, or viceversa.
\bibitem{comm3} Note that $p_{d2}(\omega)>0$ is always verified for $\omega<\omega_{\rm max}$
provided that $k_{d2-\rm{out}}(\omega_{\rm max})<q_{d}$.
\end{thebibliography}
\end{document}